\documentclass[sigconf,natbib=true]{acmart}  

\usepackage[utf8]{inputenc}
\usepackage{enumitem}
\usepackage{multirow}
\usepackage{colortbl}

\usepackage[font=small,labelfont=bf]{caption}

\newcommand{\thumbsup}[0]{\raisebox{-2pt}{\includegraphics[width=0.3cm]{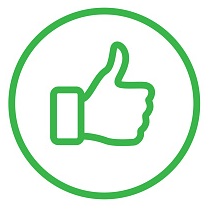}}}
\newcommand{\thumbsdown}[0]{\raisebox{-2pt}{\includegraphics[width=0.3cm]{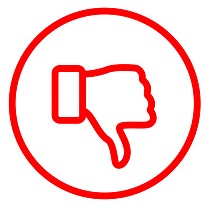}}}

\definecolor{forestgreen}{rgb}{0.13, 0.55, 0.13}
\definecolor{brickred}{rgb}{0.8, 0.25, 0.33}

\newcommand{\greenup}{\textcolor{forestgreen}{$\blacktriangle$}}
\newcommand{\reddown}{\textcolor{brickred}{$\blacktriangledown$}}

\title{Measuring the Impact of Explanation Bias: A Study of Natural Language Justifications for Recommender Systems}

\author{Krisztian Balog}
  \affiliation{%
    \institution{Google}
    \city{Stavanger}
    \country{Norway}}
  \email{krisztianb@google.com}

\author{Filip Radlinski}
  \affiliation{%
    \institution{Google}
    \city{London}
    \country{UK}}
  \email{filiprad@google.com}

\author{Andrey Petrov}
  \affiliation{%
    \institution{Google}
    \city{London}
    \country{UK}}
  \email{apetrov@google.com}

\copyrightyear{2023}
\acmYear{2023}
\setcopyright{rightsretained}
\acmConference[CHI EA '23]{Extended Abstracts of the 2023 CHI Conference on Human Factors in Computing Systems}{April 23--28, 2023}{Hamburg, Germany}
\acmBooktitle{Extended Abstracts of the 2023 CHI Conference on Human Factors in Computing Systems (CHI EA '23), April 23--28, 2023, Hamburg, Germany}\acmDOI{10.1145/3544549.3585748}
\acmISBN{978-1-4503-9422-2/23/04}

\begin{document}

\begin{abstract}
Despite the potential impact of explanations on decision making, there is a lack of research on quantifying their effect on users' choices.
This paper presents an experimental protocol for measuring the degree to which positively or negatively biased explanations can lead to users choosing suboptimal recommendations.
Key elements of this protocol include a preference elicitation stage to allow for personalizing recommendations, manual identification and extraction of item aspects from reviews, and a controlled method for introducing bias through the combination of both positive and negative aspects. We study explanations in two different textual formats: as a list of item aspects and as fluent natural language text.
Through a user study with 129 participants, we demonstrate that explanations can significantly affect users' selections and that these findings generalize across explanation formats. 
\end{abstract}



\maketitle

\section{Introduction}

Recommender systems have become pervasive in modern society, leading to many studies of the degree of trust people place in provided recommendations~\citep{Harman:2014:RecSys,Kunkel:2019:CHI}.  As a result, significant attention has been paid to equip these systems with explanation facilities to help users make informed decisions~\citep{Tintarev:2015:Book,Nunes:2017:UMUAI,Zhang:2020:FnTIR}.
In one of the earliest studies, \citet{Herlocker:2000:CSCW} showed that explanations can make it more likely that people will adopt the recommendations made.  Yet, studies measuring the eventual satisfaction with recommendations found that certain types of explanations can cause users to over- or underestimate the real value of a recommended item~\citep{Bilgic:2005:IUI,Gedikli:2014:IJHCS}. 
Even though explanations are known to be able to significantly affect the decision-making process of users~\citep{Tintarev:2015:Book}, significant gaps remain when it comes to measuring and understanding their effects~\citep{Piscopo:2022:SIGIRForum}.
The bulk of studies of explanations focus on the subjective question of how people perceive the recommendations~\cite{Balog:2020:SIGIR,Musto:2019:UMAP,Nunes:2017:UMUAI,Chang:2016:RecSys,Gedikli:2014:IJHCS}. 
Less attention has been paid to how to objectively quantify the degree to which explanations affect the choices that people make.  This is closely related to the concept of \emph{persuasiveness}, which is the ability of an explanation ``to convince the user to accept or disregard certain items''~\citep{Gedikli:2014:IJHCS}.
Past research has differentiated between over- and underestimate-oriented persuasiveness of \emph{explanation types}~\citep{Bilgic:2005:IUI,Gedikli:2014:IJHCS}. 
Instead, we study in quantitative detail how bias in the \emph{content} of natural language explanations can impact users' choices.

A key difference to past work is that prior research has focused almost exclusively on explanations that highlight positive aspects of items, i.e., endorse a selection, perhaps by drawing similarities to other items the user has previously indicated that they liked.  Explanations can also help users make decisions by highlighting why \emph{not} to select something.  We perform the first study that intentionally biases explanations towards positive or negative aspects of items in a controlled manner in the context of items where we have a prior understanding of which items users should find most relevant. 
We can thus quantify the degree to which positively or negatively biased explanations lead to users choosing suboptimal recommendations.  To achieve this, we are inspired by the work of \citet{Musto:2019:UMAP} who ``\emph{processed and analyzed the reviews in order to obtain a set of characteristics that are often discussed in the reviews (with a positive sentiment, of course) and can induce the user in enjoying the recommended item.}'' However, we identify positive as well as \emph{negative} aspects---as argued by~\citet{Bilgic:2005:IUI}, ``\emph{the goal of an explanation should not be to `sell' the user on the item but rather to help the user to make an informed judgment.}''

The primary research question we address is how to quantify the degree to which explanations influence user decisions, either in the positive or negative direction when presented with a set of recommendations from which to choose.  
Our key contribution is the design of an experiment protocol that allows for a quantitative study of the impact of explanations on users' choices.
Our second contribution is a preliminary analysis of the results obtained with 129 subjects.
We provide baseline statistics showing how often users presented with biased explanations can end up selecting less relevant suggestions. 
Beyond this, we study the effect of the \emph{format} of explanations, directly comparing two common formats, namely lists of item aspects and natural language text. We show that while these formats lead to a similar overall behavior, the effect of explanation bias is more pronounced in case of aspect lists than for natural language text.  These may be considered as factors when designing explainable recommender systems. Overall, this work provides a framework for measuring and optimizing explanations to help guide users to make informed decisions.

\section{Related work}

Research on explainable recommendation has focused on different dimensions, including (1) explanation goal 
(e.g., transparency, trust, effectiveness, persuasiveness),
(2) style (e.g., content-based, collaborative-based, knowledge/utility-based), 
(3) scope (e.g., user model, process, recommended item), and
(4) format (e.g., textual, visual)~\citep{Tintarev:2015:Book,Nunes:2017:UMUAI,Zhang:2020:FnTIR}.
Another way of categorization is by the method used to generate explanations: self-explainable recommender models vs. post-hoc explanations (i.e., \emph{justifications})~\citep{Biran:2017:IJCAI}.
Our approach can be classified as content-based, post-hoc, natural language explanation generation.
We do not aim to explain \emph{why} a given item was recommended, but rather provide the user with additional information (relevant characteristics of items) with the purpose of aiding them in their decision making. 

Various graphical and textual explanation formats have been considered in the past~\citep{Herlocker:2000:CSCW,Gedikli:2014:IJHCS}, with natural language being the most popular both historically~\citep{Nunes:2017:UMUAI} and in recent years~\citep{Balog:2020:SIGIR,Chang:2016:RecSys,Musto:2019:UMAP,Ni:2019:EMNLP,Penha:2022:CHIIR}.
Text-based explanations range from tags or keywords~\citep{Bilgic:2005:IUI,Vig:2009:IUI} to 
single or multiple sentences~\citep{Chang:2016:RecSys,Balog:2020:SIGIR,Tintarev:2012:UMUAI,Penha:2022:CHIIR,Musto:2019:UMAP}.
In this work, we consider both short descriptive phrases (aspects) and fluent natural language summaries as explanations.  Unique to our approach is that the summaries are generated directly from the aspects, allowing for a direct comparison between the two explanation formats.

Reviews have been exploited for improving recommendations, leading to a line of work on review-aware recommender systems~\citep{Chen:2015:UMUAI,Hernandez-Rubio:2019:UMUAI}, naturally endowing them with a higher degree of explainability and transparency~\citep{He:2015:CIKM}.
In contrast, our work continues the thread of research on leveraging reviews for the purpose of generating post-hoc recommendation justifications, treating the recommender engine as a black box~\citep{Muhammad:2016:IUI,Musto:2019:UMAP,Chang:2016:RecSys,Ni:2019:EMNLP,Penha:2022:CHIIR}.
\citet{Muhammad:2016:IUI} highlight the most important features (pros and cons) of an item that are likely to matter to the user (based on their own reviews).
Similarly, \citet{Chen:2014:WWW,Chen:2017:IUI} extract sentiments on specific product attributes and use them to present the user with alternatives to a given recommendation and explaining the trade-offs, i.e., which attributes would be improved and which ones would be compromised.  
\citet{Chang:2016:RecSys} employ a multi-step pipeline for generating natural language explanations using crowdsourcing, by (1) refining algorithmically generated tag clusters,  
(2) writing explanations for these clusters by synthesizing review text, and (3) selecting the best explanations by voting.  
We follow a similar approach, but present crowd workers with a significantly simpler task: given a set of reviews, they only need to extract positive and negative aspects from them.  We then either present these aspects as a list, or turn them into fluent natural language text using state-of-the-art neural language modeling techniques.
\citet{Ni:2019:EMNLP} identify review segments that can serve as justifications and explore the use of neural language models to generate convincing and diverse justifications.
\citet{Penha:2022:CHIIR} choose a helpful sentence for an item from its review, predicted by a classifier, and use that to generate a template-based explanation of an item or a pair of items.
Most closely related to ours is the work by~\citet{Musto:2019:UMAP}, which presents a fully automated pipeline for generating natural language justifications by (1) extracting a set of aspects that characterize the item, (2) identifying the most relevant ones, and (3) extracting and aggregating review sentences discussing these aspects.
Though there are similarities in the explanation generation workflow, their focus is on the algorithmic aspects of automation, while ours is on understanding the \emph{impact} of explanations.  
Our aspects are short descriptive phrases, not limited to nouns as in~\citep{Musto:2019:UMAP}, which we extract and curate manually.
Most importantly, we include aspects with negative sentiment as well, not only positive ones, in the explanations.

Evaluation of explanations may target interface-related aspects~\citep{Chen:2017:IUI,Muhammad:2016:IUI,Bilgic:2005:IUI,Herlocker:2000:CSCW,Gedikli:2014:IJHCS}, 
quantifiable properties of generation approaches (e.g., readability of natural language~\citep{Costa:2018:IUI} or fidelity of post-hoc explanations~\citep{Peake:2018:KDD}), 
or subjective perceptions of quality (in terms of transparency, trust, effectiveness, etc.) through questionnaires~\citep{Balog:2020:SIGIR,Musto:2019:UMAP,Nunes:2017:UMUAI,Chang:2016:RecSys,Gedikli:2014:IJHCS}.
Most relevant to our work is the evaluation protocol proposed in \citep{Bilgic:2005:IUI} (also followed in \citep{Gedikli:2014:IJHCS}) for measuring the \emph{persuasiveness} of explanation types, in terms of differences between an initial rating, given based on the explanation, and a second rating, given after the ``consumption'' of the item.
Consumption is approximated by subjects receiving more details (description or user reviews) about the item.  Then, they are asked adjust their rating based on this additional information.
Our experiment design is more realistic in that we do not assume item consumption, but rather measure the impact of explanations via the item selection choices users make, with a prior understanding of what they would most likely pick. 

\section{Study Design}
\label{sec:design}

This section presents the study we designed in order to answer the following two research questions:

\begin{itemize}[topsep=5pt]
    \item \textbf{RQ1} How much can explanations influence users' decisions when presented with a set of item recommendations?
    \item \textbf{RQ2} How big a role does the presentation format of explanations play in the results?
\end{itemize}
Our experiment centers around a set of item selection tasks, with participants asked to choose among three items that are presented to them as recommendations. 
These items are selected, based on an initial preference elicitation phase, such that we know which of the three items the user would probably like the most and which one the least. 
By applying a controlled amount of bias via explanations to a randomly chosen item, we are able to measure how much users may be influenced in their choices---it is of particular interest to observe how often they would be inclined to pick a different item than that which they would most likely choose (statistically speaking).
Using a within-subject design, participants are exposed to different experimental conditions, varying the amount of bias and the format of explanations.

\subsection{Overview}

The study is conducted in the movies domain.  As watching films is one of the top media consumption and entertainment activities, most people can easily relate to it without any training or prerequisite knowledge. Indeed, movies have been one of the most studied domains for explainable recommendation~\citep{Chang:2016:RecSys,Balog:2019:SIGIR,Balog:2020:SIGIR,Vig:2009:IUI,Tintarev:2012:UMUAI,Musto:2019:UMAP,Ghazimatin:2021:WWW,Herlocker:2000:CSCW,Gedikli:2014:IJHCS}.
Note that even though the instructions are tailored specifically to movies, the proposed design is generic and can be applied in other domains.

\begin{figure}[t]
    \centering
    \includegraphics[width=0.45\textwidth]{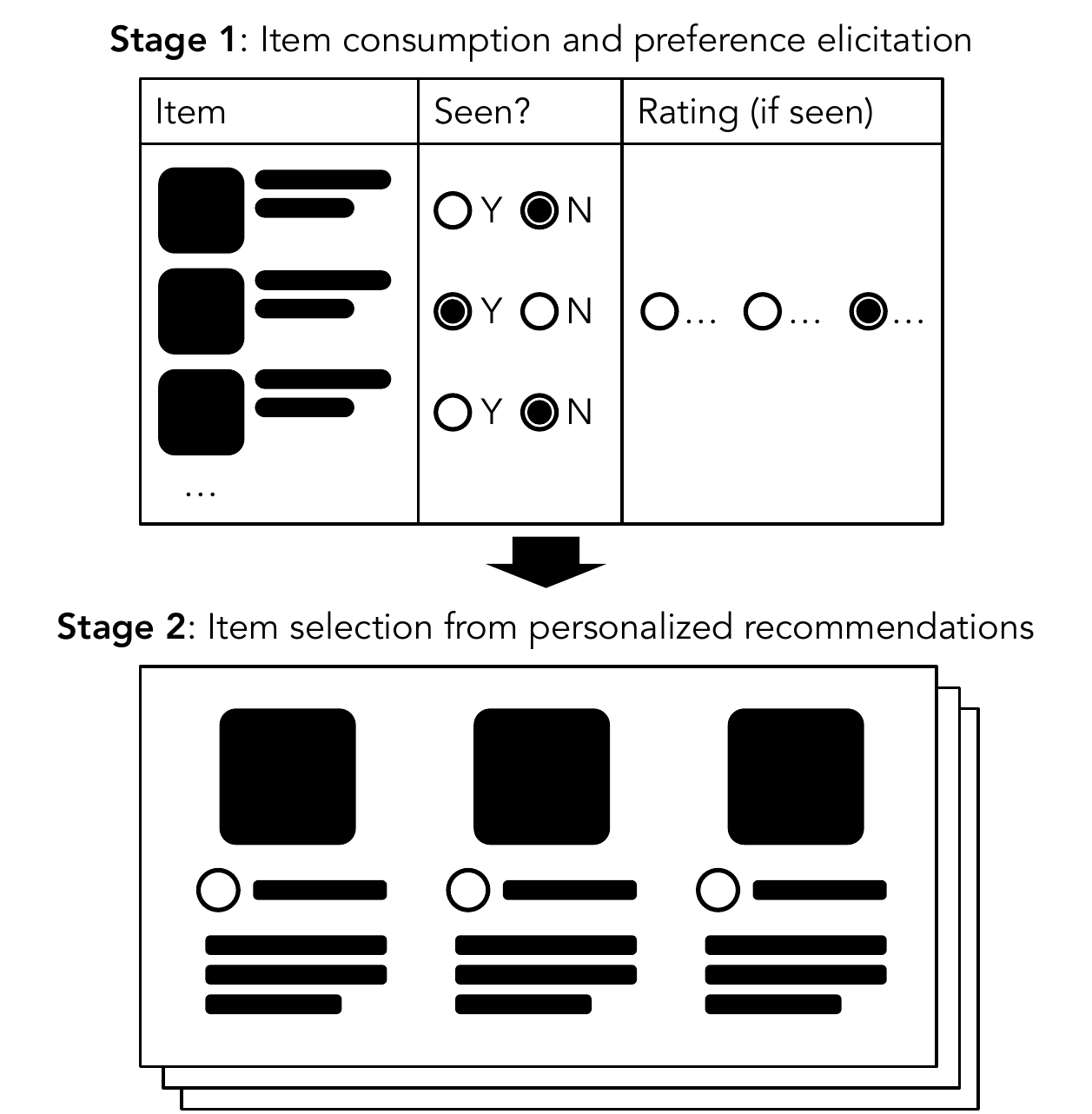}
    \vspace*{-0.5\baselineskip}
    \caption{Overview of our experiment design.}
    \vspace*{-0.5\baselineskip}
    \label{fig:overview}
    \Description{Schematic view of the two stages of the experiment design. The top figure corresponds to Stage 1 and has a 3-column tabular layout. The headings are: "Item," "Seen?," and "Rating (if seen)". Items have placeholders for image and title, seen has radio buttons with Y and N values, and rating has a number of radio buttons. The bottom figure corresponds to Stage 2, where three items are shown next to each other with image, title, description placeholders and a radio button next to each for the user to make a selection.}
\end{figure}

To be able to provide participants with personalized recommendations, the study is divided into two  stages, as shown in Fig.~\ref{fig:overview}.\footnote{Additionally, participants are asked to fill out a post-task questionnaire to allow for a qualitative analysis of their preferences in the future.}
In Stage 1, participants indicate for each movie, in our pool of 400 candidates items, whether they have watched it, and if yes, how much they liked it (on a three-point scale of disliked, neutral, and liked).
In Stage 2, participants complete a total of 12 item selection tasks, corresponding to various experimental conditions in a random order (cf.~Table~\ref{tab:conditions}).  In each task, they are offered three movies to choose from, selected among the items they have not watched yet.
For each movie the title, poster, and synopsis are shown, followed (optionally) by an explanation (see Fig.~\ref{fig:task2} in Appendix~\ref{app:ui}).
The item selection task is situated in the following scenario:
\emph{Imagine that you are traveling alone, and while in transit, you have been offered a free movie rental from a service that only allows you to choose from three possible options. In each of the following tasks, select the movie you would prefer to watch based on the information provided.}

We hypothesize that the explanations accompanying the recommendations can influence users' item selection choices.
To test this, we contrast a baseline setting, showing neutral explanations for all three items, with a ``biased'' setting: showing neutral explanations for two of the items, while making the explanation for the third item overall more positive or negative in a controlled manner.
Next, we provide details on each element of the experiment design. 

\begin{table}[t]
    \centering
    \caption{Experimental conditions. Each participant experiences each of the listed conditions exactly once (in random order).}
    \vspace*{-0.5\baselineskip}
    \label{tab:conditions}
    \begin{tabular}{ |r|l|l|c|c| } 
        \hline
        \textbf{ID} & \textbf{Bias} & \textbf{Explanation} & \textbf{Aspects$^+$} & \textbf{Aspects$^-$}  \\ 
        \hline
        \hline
        \multicolumn{5}{|l|}{\emph{Baselines}} \\    
        \hline
        \#1 & \multirow{4}{*}{No bias} & \multirow{2}{*}{-} & \multirow{2}{*}{-} & \multirow{2}{*}{-} \\
        \#2 & & & & \\
        \cline{3-5}
        \#3 & & Itemized & \multirow{2}{*}{2} & \multirow{2}{*}{2} \\
        \cline{3-3}
        \#4 & & Fluent-NL & & \\
        \hline
        \multicolumn{5}{|l|}{\emph{One random item biased by a more positive/negative explanation}} \\    
        \hline
        \#5 & \multirow{2}{*}{Bias +} & Itemized & \multirow{2}{*}{3} & \multirow{2}{*}{1} \\
        \cline{3-3}
        \#6 & & Fluent-NL & & \\
        \hline
        \#7 & \multirow{2}{*}{Bias -} & Itemized & \multirow{2}{*}{1} & \multirow{2}{*}{3} \\
        \cline{3-3}
        \#8 & & Fluent-NL & & \\
        \hline
        \#9 & \multirow{2}{*}{Bias ++} & Itemized & \multirow{2}{*}{4} & \multirow{2}{*}{0} \\
        \cline{3-3}
        \#10 & & Fluent-NL & & \\
        \hline
        \#11 & \multirow{2}{*}{Bias -{-}} & Itemized & \multirow{2}{*}{0} & \multirow{2}{*}{4} \\
        \cline{3-3}
        \#12 & & Fluent-NL & & \\
        \hline
    \end{tabular}
\end{table}

\subsection{Item Collection}
\label{sec:design:items}

We use the MovieLens 25M collection~\citep{Harper:2015:TIIS} along with the Movies and TV subset of the Amazon Reviews 2018 dataset~\citep{Ni:2019:EMNLP}.
Mappings between the two are provided by the Reviews2Movielens dataset (v2 mappings) 
released in~\citep{Zemlyanskiy:2021:EACL}.
A total of 400 movies are sampled, using a slightly modified version of the stratified sampling approach employed in~\citep{Balog:2019:SIGIR}: (i) the top 150 movies of all time by the number of ratings received and (ii) a random movie for each year between 1992 and 2016, and for each of the top 10 most popular genres (action, adventure, documentary, comedy, crime, drama, horror, romance, sci-fi, and thriller), that is not already in the top-150 set.\footnote{If it was not possible, then a random movie is sampled from the same year.}  
We require each movie to have at least 100 associated reviews.

\subsection{Generating Personalized Recommendations}
\label{sec:design:recommendations}

Study participants are presented with a set of item selection tasks.  In each task, they need to make a choice between three items $x_a$, $x_b$, and $x_c$, shown in random order.  These items are selected, based on the initial preference elicitation phase, such that the user's estimated preference ordering is $x_a \succ x_b \succ x_c$.
For that, we employ an ensemble of three established recommender algorithms, representing different classes of collaborative filtering approaches that performed best in a prior user study on movie recommendations~\citep{Balog:2019:SIGIR}:
Item-based k-Nearest Neighbors~\citep{Sarwar:2001:WWW}, 
Weighted Regularized Matrix Factorization~\citep{Hu:2008:ICDM}, 
and a Sparse Linear Method~\citep{Ning:2011:ICDM}. 
The predictions of these algorithms are combined into an ensemble recommendation using a consensus-based voting system (Borda count).  This ensemble is expected to yield better performance than any of the individual recommenders. 
Crucially, only movies not yet seen by the user are eligible for recommendation, as showing already seen items might affect the measurements of the impact of explanations~\citep{Balog:2020:SIGIR}.

\begin{figure*}
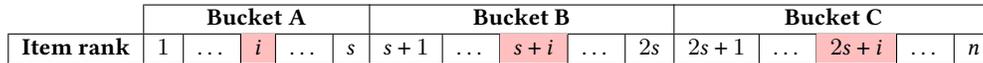

    \vspace*{-0.5\baselineskip}
    \begin{tabular}{c|c|c|c|c|c|c|c|c|c|c|c|c|c|c|c|}
        \cline{2-16}
        & \multicolumn{5}{c|}{\textbf{Bucket A}} & \multicolumn{5}{c|}{\textbf{Bucket B}} & \multicolumn{5}{c|}{\textbf{Bucket C}} \\
        \hline
        \multicolumn{1}{|l|}{\textbf{Item rank}} & 
            $1$ & $\dots$ & \cellcolor{red!25}$i$ & $\dots$ & $s$ &
            $s+1$ & $\dots$ & \cellcolor{red!25}$s+i$ & $\dots$ & $2s$ &
            $2s+1$ & $\dots$ & \cellcolor{red!25}$2s+i$ & $\dots$ & $n$ \\
        \hline
    \end{tabular}
    \vspace*{-0.5\baselineskip}
    \caption{Sampling of items for item selection tasks to ensure the largest distance between sampled items across all item samples.}
    \label{fig:item_sampling}
    \Description{Table with three columns: Bucket A, Bucket B, and Bucket C. Within each column the item ranks are shown. In Bucket A these range from 1 to s, in Bucket B from s+1 to 2s, and in Bucket C from 2s+1 to n. Item ranks i, s+i, and 2s+i within Buckets A, B, and C, respectively, are highlighted.}
\end{figure*}

\renewcommand{\arraystretch}{1.25}
\begin{figure*}[t]
    \small
    \begin{tabular}{|ll|ll|ll|}
        \hline
            \thumbsup & funny, silly, sweet & 
            \thumbsup & good, clean, and exciting movie &
            \thumbsup & suspenseful \\
            \thumbsup & awesome comedy & 
            \thumbsup & full of action &
            \thumbsdown & a bad remake \\
            \thumbsdown & a few little laughs but not enough & 
            \thumbsup & entertainment for the whole family &
            \thumbsdown & predictable \\
            \thumbsdown & lackluster direction &
            \thumbsdown & predictable attempt at comedy &
            \thumbsdown & low budget effects \\
        \hline
        \hline
        \multicolumn{2}{|p{5cm}|}{This movie is said to be funny, silly, sweet, and an awesome comedy. However, others say it has a few little laughs but not enough, and a lackluster direction.}
        &
        \multicolumn{2}{p{5cm}|}{This movie is said to be good, clean, and exciting, full of action, and an entertainment for the whole family, but some people find it to be a predictable attempt at comedy.}
        &
        \multicolumn{2}{p{5cm}|}{This movie is said to be suspenseful, but others find it to be a bad remake and predictable, with low-budget effects.} \\
        \hline
    \end{tabular}
    \vspace*{-0.5\baselineskip}
    \caption{Examples of itemized and fluent-NL explanations, based on the same set of aspects.} 
    \label{fig:explanation}
    \Description{Table with three columns and two rows. The columns illustrate different explanations, the rows correspond to different explanation styles: itemized and fluent text. Itemized explanations have green thumbs up and red thumbs down icons as the bullet symbol. The left column has two positive and two negative aspects, the middle column has three positive and one negative aspects, and the right column has one positive and three negative aspects.}
\end{figure*}
\renewcommand{\arraystretch}{1.0}

Given a total of $m=12$ item selection tasks to be completed by each participant (cf. Table~\ref{tab:conditions}), $m \times 3$ items are sampled as follows.\footnote{Participants with fewer than 3$\times$12 not yet seen movies are excluded from the study.}
Let $n$ denote the number of movies the person has not seen yet, which are sorted by recommendation score, such that $x_1$ is the highest and $x_n$ is the lowest ranked suggestion.  These items are divided into three approximately equal-sized ($s=\lfloor n/3 \rfloor$) buckets A, B, and C.
To sample $m$ sets of three items, the items picked in set $i \in [1..m]$ are: $x_a=x_i$, $x_b=x_{s+i}$, and $x_c=x_{2s+i}$. 
This ensures that the three items shown to the user are as far apart from each other as possible, in terms of recommendation score, across all item selection tasks; see Fig.~\ref{fig:item_sampling} for a visual explanation.  The $m$ sets of items ($x_a$, $x_b$, and $x_c$) are assigned randomly to the $m$ experimental conditions.

\subsection{Generating Explanations}
\label{sec:design:explanations}

\begin{table}[t]
    \caption{Controlled generation of explanation sentiment based on the number of positive and negative aspects mentioned.}
    \vspace*{-0.5\baselineskip}
    \label{tab:aspects}
    \small
    \begin{tabular}{ |l|c|c|c|c|c| } 
        \hline
        & \textbf{Extreme} & \multirow{2}{*}{\textbf{Negative}} & \multirow{2}{*}{\textbf{Neutral}} & \multirow{2}{*}{\textbf{Positive}} & \textbf{Extreme} \\
        & \textbf{Negative} & & & & \textbf{Positive} \\
        & (Bias -{-}) & (Bias -) & (No bias) & (Bias +) & (Bias ++) \\
        \hline
        \hline
        Aspects$^+$ & 0 & 1 & 2 & 3 & 4 \\
        Aspects$^-$ & 4 & 3 & 2 & 1 & 0 \\
        \hline
    \end{tabular}
    \vspace*{-0.5\baselineskip}
\end{table}

To quantify how much explanations can influence user decisions, we need a controlled way of generating explanations which are by default neutral, but can be biased to have a more positive or negative overall sentiment.
To operationally define what it takes for an explanation to be overall neutral/positive/negative we assume that for each item $M=4$ positive and $M$ negative aspects have been identified.
An aspect in this context is a short natural language text (typically 1--5 words in length) that expresses why a movie might be liked or disliked, such as ``action-packed,'' ``quite corny and unrealistic,'' or ``classic literature brilliantly realized.''  
The sentiment of an explanation can then be controlled by adjusting how many positive and negative aspects it mentions, with Aspects$^+$\,$=$\,Aspects$^-$\,$=M/2$ representing a neutral explanation (cf.~Table~\ref{tab:aspects}).
Aspects are extracted from reviews, as detailed below in Section~\ref{sec:design:explanations:aspects}.

We consider two explanation formats: itemized and fluent natural language text.
\emph{Itemized} explanations are comprised of a list of aspects with their corresponding sentiment symbolized by a thumbs up or down icon. 
Alternatively, the \emph{fluent-NL} format presents the same information as fluent natural language description; see Fig.~\ref{fig:explanation} for an illustration.
The generation of fluent-NL explanations from a list of item aspects is detailed below in Section~\ref{sec:design:explanations:fluent_nl}.

\subsubsection{Aspect Extraction}
\label{sec:design:explanations:aspects}

Positive and negative aspects are extracted manually using crowdsourcing.  Given 10 positive (4-5 stars) and 10 negative reviews (1-2 stars), workers are asked to find aspects that complete the sentence: \emph{This movie may be liked (disliked) because it is/has/contains \_\_\_}.
The goal is not to be exhaustive, but rather focus on high data quality (i.e., favor precision over recall). 
To ensure that, extracted aspects are further checked and manually filtered by authors of the paper to remove too harsh or offensive language, aspects that are too generic (e.g., excellent, terrible) or are not about movie itself (concern price, delivery, medium, etc.).  Movies with fewer than 4 positive and 4 negative aspects after filtering are removed from the recommendation pool (31 in total).
Further details are in Appendix~\ref{app:stage2:aspects}.

\subsubsection{Fluent-NL Explanation Generation}
\label{sec:design:explanations:fluent_nl}

We turn the selected aspects into fluent natural language sentences using a large language model with state-of-the-art few-shot performance (PaLM~\citep{Chowdhery:2022:arXiv}, 62b parameter model).
A separate prompt is created for each combination of Aspects$^+$ and Aspects$^-$, containing three hand-crafted training examples.

\subsection{Participant Condition Assignment}
\label{sec:design:conditions}

The study is performed through crowdsourcing via a web-based platform.  
Each participant is presented with each of the 12 experimental conditions, shown in Table~\ref{tab:conditions}, exactly once in a random order.  In fact, there are only 11 unique conditions, as \#1 and \#2 are the same, but this ``no explanations'' setting is shown twice in order to establish a robust baseline. 
The other baseline is to show neutral explanations, in two different formats (\#3--\#4).  
In the remaining conditions (\#5-\#12) one of the three items shown to the rater is randomly selected to be ``biased,'' by changing the number of positive and negative aspects in the explanation that accompanies that item.  The explanations for the other two items stay neutral, i.e., showing exactly two positive and two negative aspects.
These conditions allow us to make comparisons between different explanation types, i.e., no explanation (\#1--\#2) vs. itemized (\#3, \#5, \#7, \#9, \#11) vs. fluent-NL (\#4, \#6, \#8, \#10, \#12) explanations, and between the amount of bias, i.e., no bias (\#1--\#4) vs. moderate (\#5--\#8) vs. large (\#9--\#12) bias.

We acknowledge that listing positive aspects first, followed by negative ones, or the other way around,  
might have an impact.  
At the same time, to avoid further complicating the design by introducing yet another dimension, we control for this variable by balancing the two possible settings, i.e., positives-first and negatives-first.
Specifically, when both positive and negative aspects are displayed (\#3--\#8), we make those fully balanced for each participant as well as across all participants by cycling through pre-defined sequences in a Latin Square-like design (see Table~\ref{tab:stage2_sequences} in Appendix~\ref{app:stage2:explanations}).

\section{Results}

We now present quantitative results that show how biasing explanations that accompany recommendations directly affects users' preferences, as expressed by their item selections. 

\subsection{Participants}

A total of 129 participants participated in our study who are paid contractors, and received a standard contracted wage (complying with living wage laws in their country of employment).  All of them are US-based and native English speakers.  In terms of gender, they are 62.7\% women, 35.7\% men, 1.6\% prefer not to say.  Their distribution by age: 5.6\% 18--24, 29.4\% 25--34, 32.5\% 35--44, 19.8\% 45--54, 9.5\% 55--64, 3.2\% over 65.  Their self-reported amount of average time spent per week watching movies: 5.7\% $<$2 hours, 37.9\% 2--5 hours, 29\% 6--10 hours, 16.9\% 11-16 hours, 10.5\% $>$ 16 hours. 
Participants on average took 41 seconds to rate a batch of five movies in Stage 1 (seen/liked), and 43 seconds to choose the movie they would want to watch from the three recommendations in Stage 2.

\subsection{Effect of the Presence of Explanations}
\label{sec:results:effect}

\begin{figure}[t]
    \centering
    \includegraphics[width=0.45\textwidth]{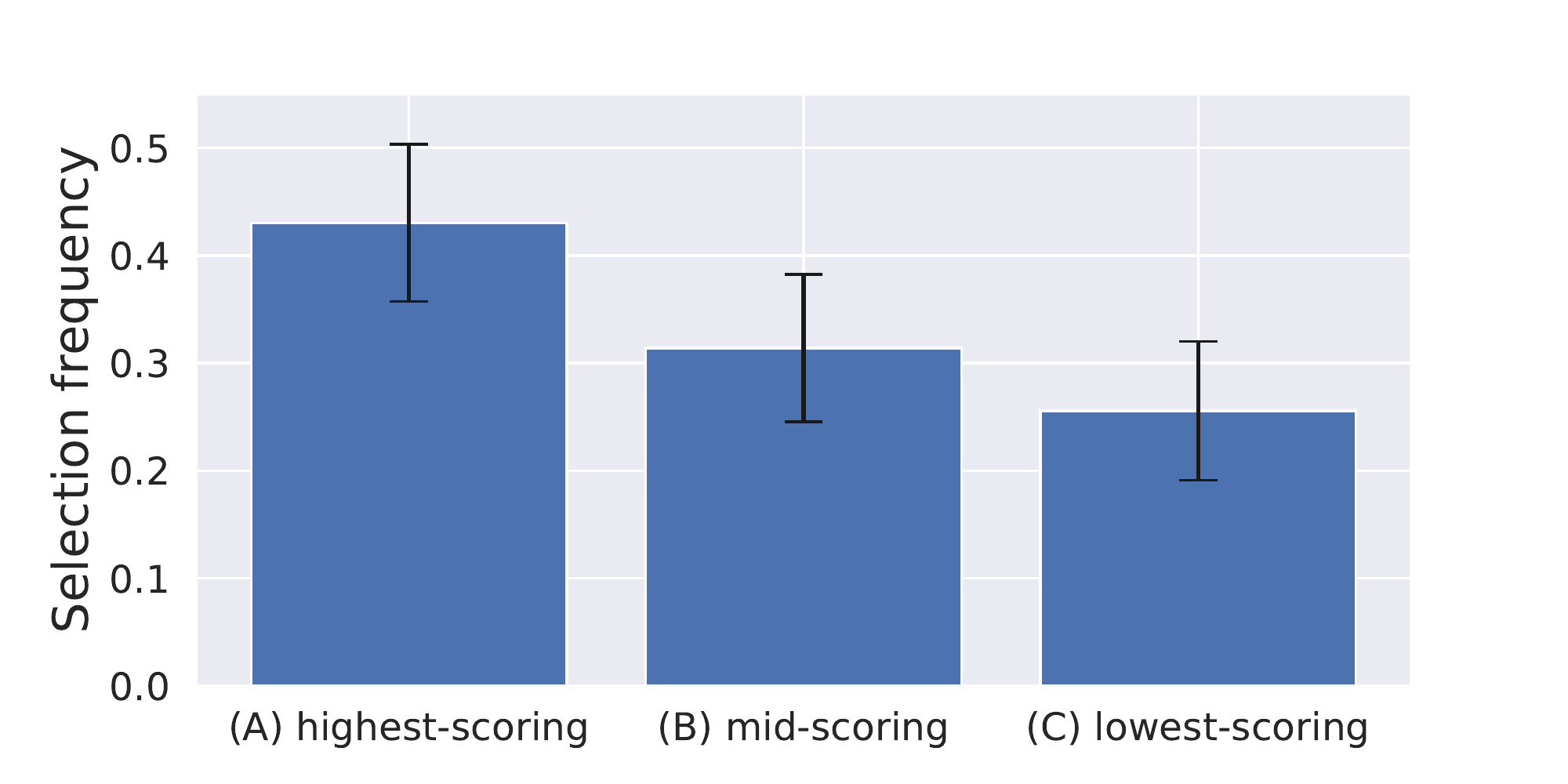}
    \vspace*{-0.5\baselineskip}
    \caption{Baseline condition, no explanations.}
    \label{fig:baseline_no_expl}
    \Description{Bar plots showing selection frequencies for (A) highest-scoring, (B) mid-scoring, and (C) lowest-scoring buckets.}
\end{figure}

Recall that in each instance, participants are provided with three recommendations where we know which the raters should prefer the most. Our first test validates that the preferences obey the expected order, and whether the presence of neutral explanations affects this order.
Specifically, let the three recommendations be $x_a \in A$, $x_b \in B$ and $x_c \in C$, where $A$, $B$, and $C$ are buckets with the highest-scoring, mid-scoring, and lowest-scoring items, respectively (cf. Fig.~\ref{fig:item_sampling}).
Let the relative frequency with which each bucket is chosen (corresponding to users' item selections), referred to as their \emph{selection frequency}, be denoted as $p_A$, $p_B$, and $p_C$ respectively.  We then have an expected ordering of $p_A \succ p_B \succ p_C$. 

When explanations are not shown, we find that across all participants, $p_A = 0.43 \pm 0.07$, $p_B = 0.31 \pm 0.07$ and $p_C = 0.26 \pm 0.06$ with a 95\% confidence interval, computed using the Goodman method~\citep{Goodman:1965:Technometrics}; see Fig.~\ref{fig:baseline_no_expl}. Thus, the expected ordering holds, and the difference between the highest- and lowest-scoring buckets is statistically significant.

\begin{figure}[t]
    \centering
    \includegraphics[width=0.45\textwidth]{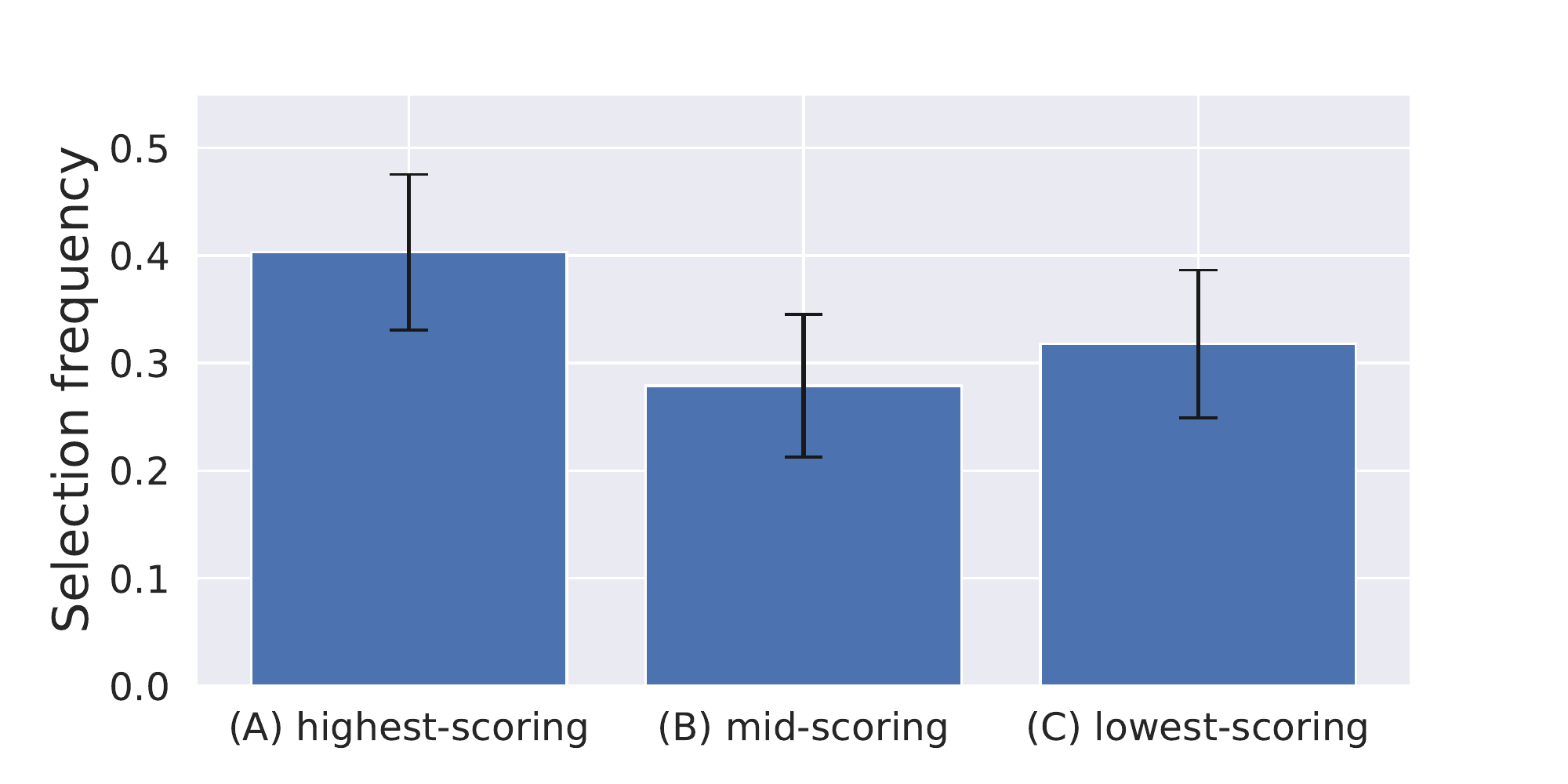}
    \vspace*{-0.5\baselineskip}
    \caption{Baseline condition, neutral explanations.}
    \label{fig:baseline_expl}
    \Description{Bar plots showing selection frequencies for (A) highest-scoring, (B) mid-scoring, and (C) lowest-scoring buckets.}
\end{figure}

When neutral explanations are included, we observe $p^n_A = 0.4 \pm 0.07$, $p^n_B = 0.28 \pm 0.07$, and $p^n_C = 0.32 \pm 0.07$, as shown in Fig.~\ref{fig:baseline_expl}.
While the highest-scoring bucket continues to receive the most selections, the relative ordering between buckets B and C, surprisingly, is now swapped.
Also, the selection frequencies for all three buckets come closer together, resulting in overlapping confidence intervals.
This means that despite the careful experiment design, explanations seem to have some uncontrolled effects.  It could be, for example, that mid- and low-scoring items are not that well distinguished by the recommender system and it is a random effect due to noise in the data.
However, the ``well-behaving'' baseline setting without explanations and the relatively large sample size (n=258) suggest otherwise.  It could also be that neutral explanations that highlight both positive and negative aspects invite more ``risky'' selections by users, giving lower-ranked suggestion a try.
In the remainder of our analysis, we take the neutral explanations setting (Fig.~\ref{fig:baseline_expl}) as our baseline.  However, the relative ordering of buckets B and C warrants further investigation.

\begin{table*}[t]
    \caption{Selection frequency of different buckets (rows) depending on the presence and direction of bias in explanations (columns).
    Grey cell background indicates when the bias happens in the same bucket as the selection.  Green/red arrows show the change in selection frequency with respect to the no bias setting.}
    \label{tab:effect_of_bias}
    \vspace*{-0.5\baselineskip}
    \centering
    \begin{tabular}{c||c||c|c||c|c||c|c}
        \hline
        \multirow{2}{*}{\textbf{Selection}} & \multirow{2}{*}{\textbf{No bias}} & \multicolumn{2}{c||}{\textbf{Bias in A}} & \multicolumn{2}{c||}{\textbf{Bias in B}} & \multicolumn{2}{c}{\textbf{Bias in C}} \\
         & & \textbf{+/++} & \textbf{-/-{-}} & \textbf{+/++} & \textbf{-/-{-}} & \textbf{+/++} & \textbf{-/-{-}} \\
        \hline
        A & 0.40\,$\pm$\,0.07 
            & \cellcolor{gray!25}0.49\,$\pm$\,0.09\greenup & \cellcolor{gray!25}0.37\,$\pm$\,0.09\reddown
            & 0.37\,$\pm$\,0.09\reddown & 0.41\,$\pm$\,0.09\greenup
            & 0.36\,$\pm$\,0.08\reddown & 0.50\,$\pm$\,0.09\greenup \\
        B & 0.28\,$\pm$\,0.07 
            & 0.25\,$\pm$\,0.08\reddown & 0.34\,$\pm$\,0.08\greenup 
            & \cellcolor{gray!25}0.42\,$\pm$\,0.09\greenup & \cellcolor{gray!25}0.30\,$\pm$\,0.08\greenup 
            & 0.27\,$\pm$\,0.08\reddown & 0.29\,$\pm$\,0.08\greenup \\ 
        C & 0.32\,$\pm$\,0.07 
            & 0.26\,$\pm$\,0.08\reddown & 0.29\,$\pm$\,0.08\reddown 
            & 0.21\,$\pm$\,0.08\reddown & 0.28\,$\pm$\,0.08\reddown 
            & \cellcolor{gray!25}0.38\,$\pm$\,0.09\greenup & \cellcolor{gray!25}0.21\,$\pm$\,0.07\reddown \\ 
        \hline
    \end{tabular}
    \vspace*{-0.5\baselineskip}
\end{table*}

\begin{table*}[t]
    \caption{Selection frequency of different buckets (rows) depending on the presence and direction of bias in explanations (columns), for itemized explanations (top block) vs. fluent-NL explanations (bottom block); cell annotations are the same as in Table~\ref{tab:effect_of_bias}.}
    \label{tab:effect_of_bias_expl_type}
    \vspace*{-0.5\baselineskip}
    \centering
    \begin{tabular}{c||c||c|c||c|c||c|c}
        \hline
        \multirow{2}{*}{\textbf{Selection}} & \multirow{2}{*}{\textbf{No bias}} & \multicolumn{2}{c||}{\textbf{Bias in A}} & \multicolumn{2}{c||}{\textbf{Bias in B}} & \multicolumn{2}{c}{\textbf{Bias in C}} \\
         & & \textbf{+/++} & \textbf{-/-{-}} & \textbf{+/++} & \textbf{-/-{-}} & \textbf{+/++} & \textbf{-/-{-}} \\
        \hline
        \multicolumn{8}{l}{\textbf{Itemized explanations}} \\
        \hline
        A & 0.40\,$\pm$\,0.10 
            & \cellcolor{gray!25}0.49\,$\pm$\,0.13\greenup & \cellcolor{gray!25}0.35\,$\pm$\,0.11\reddown
            & 0.39\,$\pm$\,0.12\reddown & 0.47\,$\pm$\,0.13\greenup
            & 0.34\,$\pm$\,0.12\reddown & 0.55\,$\pm$\,0.12\greenup \\
        B & 0.28\,$\pm$\,0.09 
            & 0.26\,$\pm$\,0.11\reddown & 0.34\,$\pm$\,0.11\greenup 
            & \cellcolor{gray!25}0.45\,$\pm$\,0.13\greenup & \cellcolor{gray!25}0.19\,$\pm$\,0.11\reddown 
            & 0.26\,$\pm$\,0.11\reddown & 0.27\,$\pm$\,0.11\reddown \\ 
        C & 0.32\,$\pm$\,0.10 
            & 0.25\,$\pm$\,0.11\reddown & 0.31\,$\pm$\,0.11\reddown 
            & 0.16\,$\pm$\,0.10\reddown & 0.33\,$\pm$\,0.12\greenup 
            & \cellcolor{gray!25}0.40\,$\pm$\,0.12\greenup & \cellcolor{gray!25}0.19\,$\pm$\,0.10\reddown \\ 
        \hline
        \multicolumn{8}{l}{\textbf{Fluent-NL explanations}} \\
        \hline
        A & 0.40\,$\pm$\,0.10 
            & \cellcolor{gray!25}0.49\,$\pm$\,0.12\greenup & \cellcolor{gray!25}0.40\,$\pm$\,0.13\phantom{\reddown}
            & 0.34\,$\pm$\,0.12\reddown & 0.36\,$\pm$\,0.12\reddown
            & 0.38\,$\pm$\,0.12\reddown & 0.45\,$\pm$\,0.12\greenup \\
        B & 0.28\,$\pm$\,0.09 
            & 0.24\,$\pm$\,0.10\reddown & 0.33\,$\pm$\,0.12\greenup 
            & \cellcolor{gray!25}0.39\,$\pm$\,0.13\greenup & \cellcolor{gray!25}0.40\,$\pm$\,0.12\greenup 
            & 0.27\,$\pm$\,0.11\reddown & 0.31\,$\pm$\,0.12\greenup \\ 
        C & 0.32\,$\pm$\,0.10 
            & 0.26\,$\pm$\,0.10\reddown & 0.27\,$\pm$\,0.12\reddown 
            & 0.27\,$\pm$\,0.12\reddown & 0.24\,$\pm$\,0.11\reddown 
            & \cellcolor{gray!25}0.35\,$\pm$\,0.12\greenup & \cellcolor{gray!25}0.23\,$\pm$\,0.11\reddown \\ 
        \hline    
    \end{tabular}
\end{table*}

\subsection{Effect of Explanations Biased towards Positive or Negative}

By adjusting the number of positive and negative aspects, we can bias explanations in a positive or negative direction either weakly (e.g., three positive aspects and one negative aspect) or strongly (e.g., four negative aspects and no positives). 
Table~\ref{tab:effect_of_bias} shows the effect of bias on item selections (rows) depending on the position and direction of bias (columns).
For simplicity, we do not distinguish between the amount of bias (weakly or strongly positive/negative) nor the type of explanation (fluent-NL or itemized), but report on aggregated counts.

Bias in a given bucket has an effect on selections both in the same bucket (highlighted as grey in Table~\ref{tab:effect_of_bias}) and in other buckets.
For example, biasing positively items in C increases selections in C, but also decreases selections in A and~B.
Similarly, negatively biasing items in A drives selections down in A, while moving selections up in B.
This intuitively makes sense, but there are a few exceptions when this expected behavior cannot be observed, e.g., negative bias in A increases selections in B, but not in C.  
Despite these anomalies that remain to be investigated in the future, it is clear that biasing explanations has a large effect on the selections people make.  Two extremes are worth noting: (1) negatively biasing the most relevant recommendation reduces the selection of what is believed to be the best recommendation by 9\%, and (2) positively biasing the least relevant recommendation increases the selection of that item by twice as much, almost 19\%, compared to the no bias baseline.
Notice that the selection frequency of the least relevant suggestion with a positive bias (0.38\,$\pm$\,0.09) reaches that of the most relevant suggestion with a negative bias (0.37\,$\pm$\,0.09).

\subsection{Effect of Explanation Format: Fluent-NL vs. Itemized Explanations}

In our design, explanations were presented in two ways: as a list of attributes, and as fluent text that mentions the same attributes. 
The last question we ask is: To what extent do our findings depend on the particular explanation format?
Table~\ref{tab:effect_of_bias_expl_type} breaks down the previous results by explanation format. Our main findings are as follows.
First, there is no difference in results in the no bias setting.
Second, when explanations are biased, itemized explanations behave more ``as expected,'' i.e., when bias happens in the same bucket as the selection (grey cells), then positive bias always means an increase and negative bias always causes a drop in selection frequency.  This is not the case for fluent-NL explanations. 
Third, we observe that the differences between the positive and negative bias settings within a given bucket tend to be much larger in case of itemized explanations.  This intuitively makes sense, as the positive and negative aspects are made explicit with visual thumbs up/down icons, while fluent text can be more prone to hiding differences.
Despite these differences, results indicate a consistent pattern of change across the two explanation formats: the arrows indicating change point in the same direction in 14 out of the 18 cells.

\section{Conclusion}

We have designed a protocol that allows for a quantitative study of the impact of explanations on users' choices in item recommendation.
Key elements of this design include preference elicitation that allows for the generation of personalized recommendations, manual identification and extraction of item aspects to include in explanations, a controlled way of introducing bias via the combination of both positive and negative aspects, and the presentation of explanations in two different textual formats.
We have conducted a user study and showed that explanations can indeed have a large effect on the item selections that people make, and that these findings generalize across the two explanation formats.
The results have also yielded some unexpected findings that warrant further investigation in future work.
We also plan to conduct a more detailed statistical analysis of the results and perform qualitative evaluation based on post-survey responses.
Further, the differences in terms of absolute impact between itemized and fluent natural language explanations suggest that the specific wording of the latter might play a role.  Measuring whether slight differences in phrasing have an impact is an interesting topic for future research.
Finally, we focused on movie recommendations, yet our approach is generalizable to other domains where users rely on automatic suggestions due to the size of the item collection (e.g., books, music, recipes). It would be interesting to repeat the experiment in other domains.

\bibliographystyle{ACM-Reference-Format}
\bibliography{references}


\begin{thebibliography}{34}


\ifx \showCODEN    \undefined \def \showCODEN     #1{\unskip}     \fi
\ifx \showDOI      \undefined \def \showDOI       #1{#1}\fi
\ifx \showISBNx    \undefined \def \showISBNx     #1{\unskip}     \fi
\ifx \showISBNxiii \undefined \def \showISBNxiii  #1{\unskip}     \fi
\ifx \showISSN     \undefined \def \showISSN      #1{\unskip}     \fi
\ifx \showLCCN     \undefined \def \showLCCN      #1{\unskip}     \fi
\ifx \shownote     \undefined \def \shownote      #1{#1}          \fi
\ifx \showarticletitle \undefined \def \showarticletitle #1{#1}   \fi
\ifx \showURL      \undefined \def \showURL       {\relax}        \fi
\providecommand\bibfield[2]{#2}
\providecommand\bibinfo[2]{#2}
\providecommand\natexlab[1]{#1}
\providecommand\showeprint[2][]{arXiv:#2}

\bibitem[Balog and Radlinski(2020)]%
        {Balog:2020:SIGIR}
\bibfield{author}{\bibinfo{person}{Krisztian Balog} {and}
  \bibinfo{person}{Filip Radlinski}.} \bibinfo{year}{2020}\natexlab{}.
\newblock \showarticletitle{Measuring Recommendation Explanation Quality: The
  Conflicting Goals of Explanations}. In \bibinfo{booktitle}{\emph{Proceedings
  of the 43rd International ACM SIGIR Conference on Research and Development in
  Information Retrieval}} \emph{(\bibinfo{series}{SIGIR '20})}.
  \bibinfo{pages}{329--338}.
\newblock


\bibitem[Balog et~al\mbox{.}(2019)]%
        {Balog:2019:SIGIR}
\bibfield{author}{\bibinfo{person}{Krisztian Balog}, \bibinfo{person}{Filip
  Radlinski}, {and} \bibinfo{person}{Shushan Arakelyan}.}
  \bibinfo{year}{2019}\natexlab{}.
\newblock \showarticletitle{Transparent, Scrutable and Explainable User Models
  for Personalized Recommendation}. In \bibinfo{booktitle}{\emph{Proceedings of
  the 42nd International ACM SIGIR Conference on Research and Development in
  Information Retrieval}} \emph{(\bibinfo{series}{SIGIR'19})}.
  \bibinfo{pages}{265--274}.
\newblock


\bibitem[Bilgic and Mooney(2005)]%
        {Bilgic:2005:IUI}
\bibfield{author}{\bibinfo{person}{Mustafa Bilgic} {and}
  \bibinfo{person}{Raymond~J. Mooney}.} \bibinfo{year}{2005}\natexlab{}.
\newblock \showarticletitle{Explaining Recommendations: Satisfaction vs.
  Promotion}. In \bibinfo{booktitle}{\emph{Proceedings of Beyond
  Personalization 2005: A Workshop on the Next Stage of Recommender Systems
  Research at the 2005 International Conference on Intelligent User
  Interfaces}}. \bibinfo{pages}{13--18}.
\newblock


\bibitem[Biran and Cotton(2017)]%
        {Biran:2017:IJCAI}
\bibfield{author}{\bibinfo{person}{Or Biran} {and}
  \bibinfo{person}{Courtenay~V. Cotton}.} \bibinfo{year}{2017}\natexlab{}.
\newblock \showarticletitle{Explanation and Justification in Machine Learning:
  A Survey}. In \bibinfo{booktitle}{\emph{IJCAI 2017 Workshop on Explainable
  Artificial Intelligence}}.
\newblock


\bibitem[Chang et~al\mbox{.}(2016)]%
        {Chang:2016:RecSys}
\bibfield{author}{\bibinfo{person}{Shuo Chang}, \bibinfo{person}{F.~Maxwell
  Harper}, {and} \bibinfo{person}{Loren~Gilbert Terveen}.}
  \bibinfo{year}{2016}\natexlab{}.
\newblock \showarticletitle{Crowd-Based Personalized Natural Language
  Explanations for Recommendations}. In \bibinfo{booktitle}{\emph{Proceedings
  of the 10th ACM Conference on Recommender Systems}}
  \emph{(\bibinfo{series}{RecSys '16})}. \bibinfo{pages}{175--182}.
\newblock


\bibitem[Chen et~al\mbox{.}(2015)]%
        {Chen:2015:UMUAI}
\bibfield{author}{\bibinfo{person}{Li Chen}, \bibinfo{person}{Guanliang Chen},
  {and} \bibinfo{person}{Feng Wang}.} \bibinfo{year}{2015}\natexlab{}.
\newblock \showarticletitle{Recommender Systems Based on User Reviews: The
  State of the Art}.
\newblock \bibinfo{journal}{\emph{User Model. User-Adapt. Interact.}}
  \bibinfo{volume}{25}, \bibinfo{number}{2} (\bibinfo{date}{jun}
  \bibinfo{year}{2015}), \bibinfo{pages}{99--154}.
\newblock


\bibitem[Chen and Wang(2014)]%
        {Chen:2014:WWW}
\bibfield{author}{\bibinfo{person}{Li Chen} {and} \bibinfo{person}{Feng Wang}.}
  \bibinfo{year}{2014}\natexlab{}.
\newblock \showarticletitle{Sentiment-Enhanced Explanation of Product
  Recommendations}. In \bibinfo{booktitle}{\emph{Proceedings of the 23rd
  International Conference on World Wide Web}} \emph{(\bibinfo{series}{WWW
  '14})}. \bibinfo{pages}{239--240}.
\newblock


\bibitem[Chen and Wang(2017)]%
        {Chen:2017:IUI}
\bibfield{author}{\bibinfo{person}{Li Chen} {and} \bibinfo{person}{Feng Wang}.}
  \bibinfo{year}{2017}\natexlab{}.
\newblock \showarticletitle{Explaining Recommendations Based on Feature
  Sentiments in Product Reviews}. In \bibinfo{booktitle}{\emph{Proceedings of
  the 22nd International Conference on Intelligent User Interfaces}}
  \emph{(\bibinfo{series}{IUI '17})}. \bibinfo{pages}{17--28}.
\newblock


\bibitem[Chowdhery et~al\mbox{.}(2022)]%
        {Chowdhery:2022:arXiv}
\bibfield{author}{\bibinfo{person}{Aakanksha Chowdhery},
  \bibinfo{person}{Sharan Narang}, \bibinfo{person}{Jacob Devlin},
  \bibinfo{person}{Maarten Bosma}, \bibinfo{person}{Gaurav Mishra},
  \bibinfo{person}{Adam Roberts}, \bibinfo{person}{Paul Barham},
  \bibinfo{person}{Hyung~Won Chung}, \bibinfo{person}{Charles Sutton},
  \bibinfo{person}{Sebastian Gehrmann}, \bibinfo{person}{Parker Schuh},
  \bibinfo{person}{Kensen Shi}, \bibinfo{person}{Sasha Tsvyashchenko},
  \bibinfo{person}{Joshua Maynez}, \bibinfo{person}{Abhishek Rao},
  \bibinfo{person}{Parker Barnes}, \bibinfo{person}{Yi Tay},
  \bibinfo{person}{Noam Shazeer}, \bibinfo{person}{Vinodkumar Prabhakaran},
  \bibinfo{person}{Emily Reif}, \bibinfo{person}{Nan Du}, \bibinfo{person}{Ben
  Hutchinson}, \bibinfo{person}{Reiner Pope}, \bibinfo{person}{James Bradbury},
  \bibinfo{person}{Jacob Austin}, \bibinfo{person}{Michael Isard},
  \bibinfo{person}{Guy Gur-Ari}, \bibinfo{person}{Pengcheng Yin},
  \bibinfo{person}{Toju Duke}, \bibinfo{person}{Anselm Levskaya},
  \bibinfo{person}{Sanjay Ghemawat}, \bibinfo{person}{Sunipa Dev},
  \bibinfo{person}{Henryk Michalewski}, \bibinfo{person}{Xavier Garcia},
  \bibinfo{person}{Vedant Misra}, \bibinfo{person}{Kevin Robinson},
  \bibinfo{person}{Liam Fedus}, \bibinfo{person}{Denny Zhou},
  \bibinfo{person}{Daphne Ippolito}, \bibinfo{person}{David Luan},
  \bibinfo{person}{Hyeontaek Lim}, \bibinfo{person}{Barret Zoph},
  \bibinfo{person}{Alexander Spiridonov}, \bibinfo{person}{Ryan Sepassi},
  \bibinfo{person}{David Dohan}, \bibinfo{person}{Shivani Agrawal},
  \bibinfo{person}{Mark Omernick}, \bibinfo{person}{Andrew~M. Dai},
  \bibinfo{person}{Thanumalayan~Sankaranarayana Pillai}, \bibinfo{person}{Marie
  Pellat}, \bibinfo{person}{Aitor Lewkowycz}, \bibinfo{person}{Erica Moreira},
  \bibinfo{person}{Rewon Child}, \bibinfo{person}{Oleksandr Polozov},
  \bibinfo{person}{Katherine Lee}, \bibinfo{person}{Zongwei Zhou},
  \bibinfo{person}{Xuezhi Wang}, \bibinfo{person}{Brennan Saeta},
  \bibinfo{person}{Mark Diaz}, \bibinfo{person}{Orhan Firat},
  \bibinfo{person}{Michele Catasta}, \bibinfo{person}{Jason Wei},
  \bibinfo{person}{Kathy Meier-Hellstern}, \bibinfo{person}{Douglas Eck},
  \bibinfo{person}{Jeff Dean}, \bibinfo{person}{Slav Petrov}, {and}
  \bibinfo{person}{Noah Fiedel}.} \bibinfo{year}{2022}\natexlab{}.
\newblock \bibinfo{title}{PaLM: Scaling Language Modeling with Pathways}.
\newblock
\newblock
\showeprint[arxiv]{2204.02311}~[cs.CL]


\bibitem[Costa et~al\mbox{.}(2018)]%
        {Costa:2018:IUI}
\bibfield{author}{\bibinfo{person}{Felipe Costa}, \bibinfo{person}{Sixun
  Ouyang}, \bibinfo{person}{Peter Dolog}, {and} \bibinfo{person}{Aonghus
  Lawlor}.} \bibinfo{year}{2018}\natexlab{}.
\newblock \showarticletitle{Automatic Generation of Natural Language
  Explanations}. In \bibinfo{booktitle}{\emph{Proceedings of the 23rd
  International Conference on Intelligent User Interfaces Companion}}
  \emph{(\bibinfo{series}{IUI '18})}.
\newblock


\bibitem[Gedikli et~al\mbox{.}(2014)]%
        {Gedikli:2014:IJHCS}
\bibfield{author}{\bibinfo{person}{Fatih Gedikli}, \bibinfo{person}{Dietmar
  Jannach}, {and} \bibinfo{person}{Mouzhi Ge}.}
  \bibinfo{year}{2014}\natexlab{}.
\newblock \showarticletitle{How Should I Explain? A Comparison of Different
  Explanation Types for Recommender Systems}.
\newblock \bibinfo{journal}{\emph{Int. J. Hum. Comput. Stud.}}
  \bibinfo{volume}{72}, \bibinfo{number}{4} (\bibinfo{year}{2014}),
  \bibinfo{pages}{367--382}.
\newblock


\bibitem[Ghazimatin et~al\mbox{.}(2021)]%
        {Ghazimatin:2021:WWW}
\bibfield{author}{\bibinfo{person}{Azin Ghazimatin}, \bibinfo{person}{Soumajit
  Pramanik}, \bibinfo{person}{Rishiraj Saha~Roy}, {and}
  \bibinfo{person}{Gerhard Weikum}.} \bibinfo{year}{2021}\natexlab{}.
\newblock \showarticletitle{ELIXIR: Learning from User Feedback on Explanations
  to Improve Recommender Models}. In \bibinfo{booktitle}{\emph{Proceedings of
  the Web Conference 2021}} \emph{(\bibinfo{series}{WWW '21})}.
  \bibinfo{pages}{3850--3860}.
\newblock


\bibitem[Goodman(1965)]%
        {Goodman:1965:Technometrics}
\bibfield{author}{\bibinfo{person}{Leo~A. Goodman}.}
  \bibinfo{year}{1965}\natexlab{}.
\newblock \showarticletitle{On Simultaneous Confidence Intervals for
  Multinomial Proportions}.
\newblock \bibinfo{journal}{\emph{Technometrics}} \bibinfo{volume}{7},
  \bibinfo{number}{2} (\bibinfo{date}{may} \bibinfo{year}{1965}),
  \bibinfo{pages}{247--254}.
\newblock


\bibitem[Harman et~al\mbox{.}(2014)]%
        {Harman:2014:RecSys}
\bibfield{author}{\bibinfo{person}{Jason~L. Harman}, \bibinfo{person}{John
  O'Donovan}, \bibinfo{person}{Tarek Abdelzaher}, {and}
  \bibinfo{person}{Cleotilde Gonzalez}.} \bibinfo{year}{2014}\natexlab{}.
\newblock \showarticletitle{Dynamics of Human Trust in Recommender Systems}. In
  \bibinfo{booktitle}{\emph{Proceedings of the 8th ACM Conference on
  Recommender Systems}} \emph{(\bibinfo{series}{RecSys '14})}.
  \bibinfo{pages}{305--308}.
\newblock


\bibitem[Harper and Konstan(2015)]%
        {Harper:2015:TIIS}
\bibfield{author}{\bibinfo{person}{F.~Maxwell Harper} {and}
  \bibinfo{person}{Joseph~A. Konstan}.} \bibinfo{year}{2015}\natexlab{}.
\newblock \showarticletitle{The MovieLens Datasets: History and Context}.
\newblock \bibinfo{journal}{\emph{ACM Trans. Interact. Intell. Syst.}}
  \bibinfo{volume}{5}, \bibinfo{number}{4}, Article \bibinfo{articleno}{19}
  (\bibinfo{date}{dec} \bibinfo{year}{2015}), \bibinfo{numpages}{19}~pages.
\newblock


\bibitem[He et~al\mbox{.}(2015)]%
        {He:2015:CIKM}
\bibfield{author}{\bibinfo{person}{Xiangnan He}, \bibinfo{person}{Tao Chen},
  \bibinfo{person}{Min-Yen Kan}, {and} \bibinfo{person}{Xiao Chen}.}
  \bibinfo{year}{2015}\natexlab{}.
\newblock \showarticletitle{TriRank: Review-Aware Explainable Recommendation by
  Modeling Aspects}. In \bibinfo{booktitle}{\emph{Proceedings of the 24th ACM
  International on Conference on Information and Knowledge Management}}
  \emph{(\bibinfo{series}{CIKM '15})}. \bibinfo{pages}{1661--1670}.
\newblock


\bibitem[Herlocker et~al\mbox{.}(2000)]%
        {Herlocker:2000:CSCW}
\bibfield{author}{\bibinfo{person}{Jonathan~L. Herlocker},
  \bibinfo{person}{Joseph~A. Konstan}, {and} \bibinfo{person}{John Riedl}.}
  \bibinfo{year}{2000}\natexlab{}.
\newblock \showarticletitle{Explaining Collaborative Filtering
  Recommendations}. In \bibinfo{booktitle}{\emph{Proceedings of the 2000 ACM
  Conference on Computer Supported Cooperative Work}}
  \emph{(\bibinfo{series}{CSCW '00})}. \bibinfo{pages}{241--250}.
\newblock


\bibitem[Hernández-Rubio et~al\mbox{.}(2019)]%
        {Hernandez-Rubio:2019:UMUAI}
\bibfield{author}{\bibinfo{person}{María Hernández-Rubio},
  \bibinfo{person}{Iván Cantador}, {and} \bibinfo{person}{Alejandro
  Bellogín}.} \bibinfo{year}{2019}\natexlab{}.
\newblock \showarticletitle{A Comparative Analysis of Recommender Systems Based
  on Item Aspect Opinions Extracted from User Reviews}.
\newblock \bibinfo{journal}{\emph{User Model. User-Adapt. Interact.}}
  \bibinfo{volume}{29} (\bibinfo{date}{apr} \bibinfo{year}{2019}),
  \bibinfo{pages}{381--441}.
\newblock


\bibitem[Hu et~al\mbox{.}(2008)]%
        {Hu:2008:ICDM}
\bibfield{author}{\bibinfo{person}{Yifan Hu}, \bibinfo{person}{Yehuda Koren},
  {and} \bibinfo{person}{Chris Volinsky}.} \bibinfo{year}{2008}\natexlab{}.
\newblock \showarticletitle{Collaborative Filtering for Implicit Feedback
  Datasets}. In \bibinfo{booktitle}{\emph{Proceedings of the 2008 Eighth IEEE
  International Conference on Data Mining}} \emph{(\bibinfo{series}{ICDM
  '08})}. \bibinfo{pages}{263--272}.
\newblock


\bibitem[Kunkel et~al\mbox{.}(2019)]%
        {Kunkel:2019:CHI}
\bibfield{author}{\bibinfo{person}{Johannes Kunkel}, \bibinfo{person}{Tim
  Donkers}, \bibinfo{person}{Lisa Michael}, \bibinfo{person}{Catalin-Mihai
  Barbu}, {and} \bibinfo{person}{J\"{u}rgen Ziegler}.}
  \bibinfo{year}{2019}\natexlab{}.
\newblock \showarticletitle{Let Me Explain: Impact of Personal and Impersonal
  Explanations on Trust in Recommender Systems}. In
  \bibinfo{booktitle}{\emph{Proceedings of the 2019 CHI Conference on Human
  Factors in Computing Systems}} \emph{(\bibinfo{series}{CHI '19})}.
  \bibinfo{pages}{1--12}.
\newblock


\bibitem[Muhammad et~al\mbox{.}(2016)]%
        {Muhammad:2016:IUI}
\bibfield{author}{\bibinfo{person}{Khalil~Ibrahim Muhammad},
  \bibinfo{person}{Aonghus Lawlor}, {and} \bibinfo{person}{Barry Smyth}.}
  \bibinfo{year}{2016}\natexlab{}.
\newblock \showarticletitle{A Live-User Study of Opinionated Explanations for
  Recommender Systems}. In \bibinfo{booktitle}{\emph{Proceedings of the 21st
  International Conference on Intelligent User Interfaces}}
  \emph{(\bibinfo{series}{IUI '16})}. \bibinfo{pages}{256--260}.
\newblock


\bibitem[Musto et~al\mbox{.}(2019)]%
        {Musto:2019:UMAP}
\bibfield{author}{\bibinfo{person}{Cataldo Musto}, \bibinfo{person}{Pasquale
  Lops}, \bibinfo{person}{Marco de Gemmis}, {and} \bibinfo{person}{Giovanni
  Semeraro}.} \bibinfo{year}{2019}\natexlab{}.
\newblock \showarticletitle{Justifying Recommendations through Aspect-Based
  Sentiment Analysis of Users Reviews}. In
  \bibinfo{booktitle}{\emph{Proceedings of the 27th ACM Conference on User
  Modeling, Adaptation and Personalization}} \emph{(\bibinfo{series}{UMAP
  '19})}. \bibinfo{pages}{4--12}.
\newblock


\bibitem[Ni et~al\mbox{.}(2019)]%
        {Ni:2019:EMNLP}
\bibfield{author}{\bibinfo{person}{Jianmo Ni}, \bibinfo{person}{Jiacheng Li},
  {and} \bibinfo{person}{Julian McAuley}.} \bibinfo{year}{2019}\natexlab{}.
\newblock \showarticletitle{Justifying Recommendations using Distantly-Labeled
  Reviews and Fine-Grained Aspects}. In \bibinfo{booktitle}{\emph{Proceedings
  of the 2019 Conference on Empirical Methods in Natural Language Processing
  and the 9th International Joint Conference on Natural Language Processing}}.
  \bibinfo{pages}{188--197}.
\newblock


\bibitem[Ning and Karypis(2011)]%
        {Ning:2011:ICDM}
\bibfield{author}{\bibinfo{person}{Xia Ning} {and} \bibinfo{person}{George
  Karypis}.} \bibinfo{year}{2011}\natexlab{}.
\newblock \showarticletitle{SLIM: Sparse Linear Methods for Top-N Recommender
  Systems}. In \bibinfo{booktitle}{\emph{Proceedings of the 2011 IEEE 11th
  International Conference on Data Mining}} \emph{(\bibinfo{series}{ICDM
  '11})}. \bibinfo{pages}{497--506}.
\newblock


\bibitem[Nunes and Jannach(2017)]%
        {Nunes:2017:UMUAI}
\bibfield{author}{\bibinfo{person}{Ingrid Nunes} {and} \bibinfo{person}{Dietmar
  Jannach}.} \bibinfo{year}{2017}\natexlab{}.
\newblock \showarticletitle{A Systematic Review and Taxonomy of Explanations in
  Decision Support and Recommender Systems}.
\newblock \bibinfo{journal}{\emph{User Model. User-adapt. Interact.}}
  \bibinfo{volume}{27}, \bibinfo{number}{3-5} (\bibinfo{year}{2017}),
  \bibinfo{pages}{393--444}.
\newblock


\bibitem[Peake and Wang(2018)]%
        {Peake:2018:KDD}
\bibfield{author}{\bibinfo{person}{Georgina Peake} {and} \bibinfo{person}{Jun
  Wang}.} \bibinfo{year}{2018}\natexlab{}.
\newblock \showarticletitle{Explanation Mining: Post Hoc Interpretability of
  Latent Factor Models for Recommendation Systems}. In
  \bibinfo{booktitle}{\emph{Proceedings of the 24th ACM SIGKDD International
  Conference on Knowledge Discovery \& Data Mining}}
  \emph{(\bibinfo{series}{KDD '18})}. \bibinfo{pages}{2060--2069}.
\newblock


\bibitem[Penha et~al\mbox{.}(2022)]%
        {Penha:2022:CHIIR}
\bibfield{author}{\bibinfo{person}{Gustavo Penha}, \bibinfo{person}{Eyal
  Krikon}, {and} \bibinfo{person}{Vanessa Murdock}.}
  \bibinfo{year}{2022}\natexlab{}.
\newblock \showarticletitle{Pairwise Review-Based Explanations for Voice
  Product Search}. In \bibinfo{booktitle}{\emph{ACM SIGIR Conference on Human
  Information Interaction and Retrieval}} \emph{(\bibinfo{series}{CHIIR '22})}.
  \bibinfo{pages}{300--304}.
\newblock


\bibitem[Piscopo et~al\mbox{.}(2022)]%
        {Piscopo:2022:SIGIRForum}
\bibfield{author}{\bibinfo{person}{Alessandro Piscopo}, \bibinfo{person}{Oana
  Inel}, \bibinfo{person}{Sanne Vrijenhoek}, \bibinfo{person}{Martijn
  Millecamp}, {and} \bibinfo{person}{Krisztian Balog}.}
  \bibinfo{year}{2022}\natexlab{}.
\newblock \showarticletitle{Report on the 1st Workshop on Measuring the Quality
  of Explanations in Recommender Systems (QUARE 2022) at SIGIR 2022}.
\newblock \bibinfo{journal}{\emph{SIGIR Forum}} \bibinfo{volume}{56},
  \bibinfo{number}{2} (\bibinfo{date}{dec} \bibinfo{year}{2022}).
\newblock


\bibitem[Sarwar et~al\mbox{.}(2001)]%
        {Sarwar:2001:WWW}
\bibfield{author}{\bibinfo{person}{Badrul Sarwar}, \bibinfo{person}{George
  Karypis}, \bibinfo{person}{Joseph Konstan}, {and} \bibinfo{person}{John
  Riedl}.} \bibinfo{year}{2001}\natexlab{}.
\newblock \showarticletitle{Item-based Collaborative Filtering Recommendation
  Algorithms}. In \bibinfo{booktitle}{\emph{Proceedings of the 10th
  International Conference on World Wide Web}} \emph{(\bibinfo{series}{WWW
  '01})}. \bibinfo{pages}{285--295}.
\newblock


\bibitem[Tintarev and Masthoff(2012)]%
        {Tintarev:2012:UMUAI}
\bibfield{author}{\bibinfo{person}{Nava Tintarev} {and} \bibinfo{person}{Judith
  Masthoff}.} \bibinfo{year}{2012}\natexlab{}.
\newblock \showarticletitle{Evaluating the Effectiveness of Explanations for
  Recommender Systems}.
\newblock \bibinfo{journal}{\emph{User Model. User-Adapt. Interact.}}
  \bibinfo{volume}{22} (\bibinfo{date}{oct} \bibinfo{year}{2012}),
  \bibinfo{pages}{399--439}.
\newblock


\bibitem[Tintarev and Masthoff(2015)]%
        {Tintarev:2015:Book}
\bibfield{author}{\bibinfo{person}{Nava Tintarev} {and} \bibinfo{person}{Judith
  Masthoff}.} \bibinfo{year}{2015}\natexlab{}.
\newblock \showarticletitle{Explaining Recommendations: {D}esign and
  Evaluation}.
\newblock In \bibinfo{booktitle}{\emph{Recommender Systems Handbook}
  (\bibinfo{edition}{2nd} ed.)}, \bibfield{editor}{\bibinfo{person}{Francesco
  Ricci}, \bibinfo{person}{Lior Rokach}, \bibinfo{person}{Bracha Shapira},
  {and} \bibinfo{person}{Paul~B. Kantor}} (Eds.). \bibinfo{publisher}{Springer
  US}, Chapter~10, \bibinfo{pages}{353--382}.
\newblock


\bibitem[Vig et~al\mbox{.}(2009)]%
        {Vig:2009:IUI}
\bibfield{author}{\bibinfo{person}{Jesse Vig}, \bibinfo{person}{Shilad Sen},
  {and} \bibinfo{person}{John Riedl}.} \bibinfo{year}{2009}\natexlab{}.
\newblock \showarticletitle{Tagsplanations: Explaining Recommendations Using
  Tags}. In \bibinfo{booktitle}{\emph{Proceedings of the 14th International
  Conference on Intelligent User Interfaces}} \emph{(\bibinfo{series}{IUI
  '09})}. \bibinfo{pages}{47--56}.
\newblock


\bibitem[Zemlyanskiy et~al\mbox{.}(2021)]%
        {Zemlyanskiy:2021:EACL}
\bibfield{author}{\bibinfo{person}{Yury Zemlyanskiy}, \bibinfo{person}{Sudeep
  Gandhe}, \bibinfo{person}{Ruining He}, \bibinfo{person}{Bhargav Kanagal},
  \bibinfo{person}{Anirudh Ravula}, \bibinfo{person}{Juraj Gottweis},
  \bibinfo{person}{Fei Sha}, {and} \bibinfo{person}{Ilya Eckstein}.}
  \bibinfo{year}{2021}\natexlab{}.
\newblock \showarticletitle{{DOCENT}: Learning Self-Supervised Entity
  Representations from Large Document Collections}. In
  \bibinfo{booktitle}{\emph{Proceedings of the 16th Conference of the European
  Chapter of the Association for Computational Linguistics: Main Volume}}
  \emph{(\bibinfo{series}{EACL '21})}. \bibinfo{pages}{2540--2549}.
\newblock


\bibitem[Zhang and Chen(2020)]%
        {Zhang:2020:FnTIR}
\bibfield{author}{\bibinfo{person}{Yongfeng Zhang} {and} \bibinfo{person}{Xu
  Chen}.} \bibinfo{year}{2020}\natexlab{}.
\newblock \showarticletitle{Explainable Recommendation: A Survey and New
  Perspectives}.
\newblock \bibinfo{journal}{\emph{Found. Trends Inf. Retr.}}
  \bibinfo{volume}{14}, \bibinfo{number}{1} (\bibinfo{year}{2020}),
  \bibinfo{pages}{1--101}.
\newblock


\end{thebibliography}

\appendix
\section{Appendix: Study Design}
\label{app:stage2}

We provide further details on the study design and data preparation.

\subsection{Aspect Extraction}
\label{app:stage2:aspects}

The data collected using crowdsourcing has been automatically cleaned and pre-filtered; this includes fixing capitalization, removing trailing whitespace, filtering too long aspects, and removing aspects that are substrings of other aspects.
However, the data needed further manual cleaning and filtering by the paper authors, as not all extracted aspects fit the template (i.e., human workers did not follow the instructions closely enough), aspects may be too harsh or offensive, or sound too personal to be used as explanations.  As part of the manual cleaning process, some aspects were slightly rewritten and near-duplicates were removed. Of the 9,635 movie-aspect pairs collected originally, 7,414 remained after automatic pre-filtering, and 5,948 after the end of the manual filtering.

It is worth emphasizing that the recommendations are personalized, while the explanations accompanying them are not, i.e., all participants receiving the same recommendation under the same experimental condition will see the same explanation for that item, to ensure that there is no uncontrolled bias.

\begin{table}[t]
    \caption{Sequences determining whether to list positive or negative aspects first in explanations.}
    \label{tab:stage2_sequences}
    \vspace*{-0.5\baselineskip}
    \begin{tabular}{|c|c|c|c|c|c|c|}
        \hline
        \multirow{2}{*}{\textbf{Sequence}} & \multicolumn{6}{c|}{\textbf{Condition}} \\
        \cline{2-7}
        & \textbf{\#3} & \textbf{\#4} & \textbf{\#5} & \textbf{\#6} & \textbf{\#7} & \textbf{\#8} \\
        \hline
        \hline
        1 & N & P & P & N & N & P \\
        \hline
        2 & P & N & N & P & P & N \\
        \hline
        3 & N & P & P & N & N & P \\
        \hline
        4 & P & N & N & P & P & N \\
        \hline
        5 & N & P & P & N & N & P \\
        \hline
        6 & P & N & N & P & P & N \\
        \hline
    \end{tabular}
\end{table}

\begin{figure*}[t]
    \centering
    \includegraphics[width=0.95\textwidth]{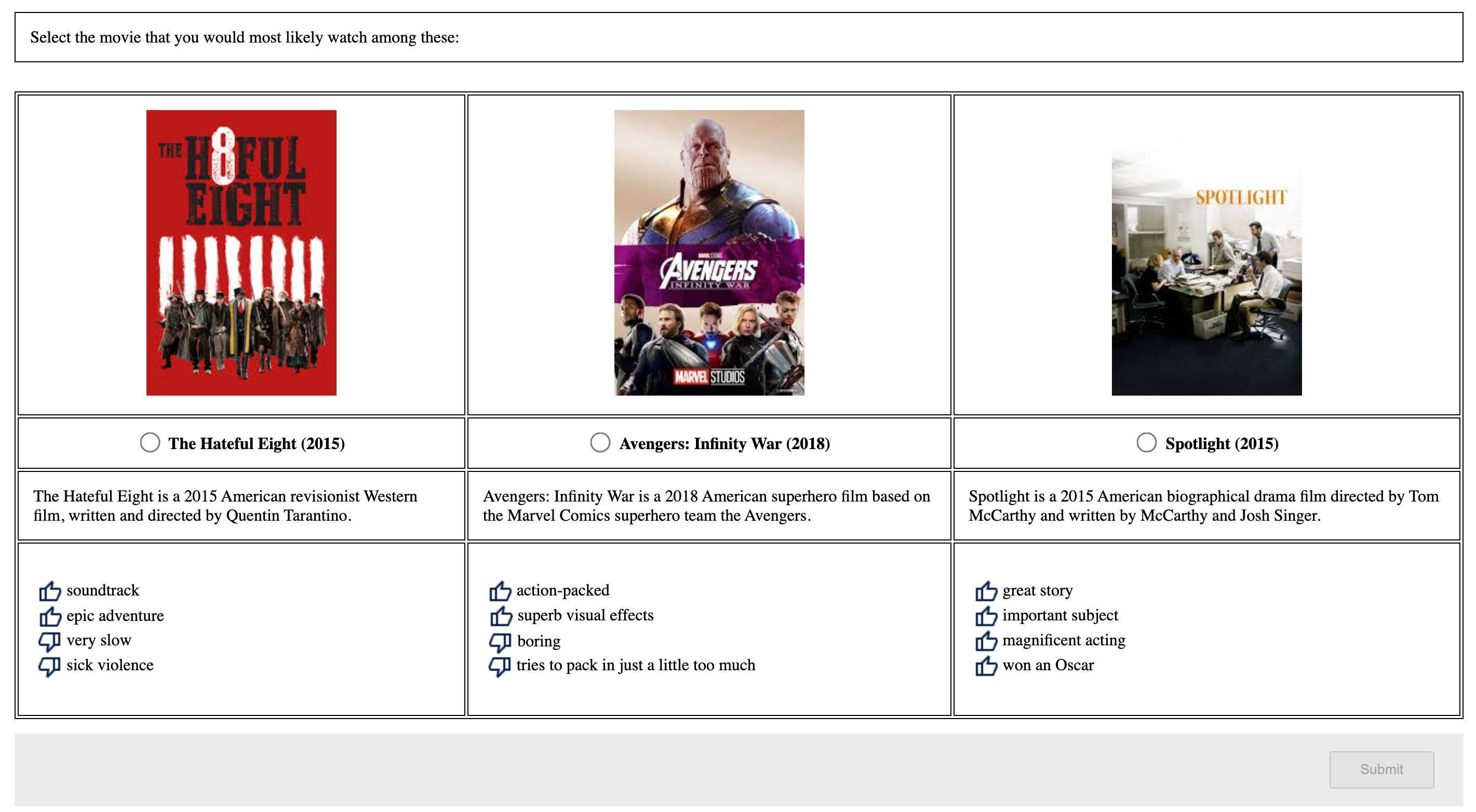}
    \vspace*{-0.5\baselineskip}
    \caption{User interface for item selection (Stage 2).}
    \label{fig:task2}
    \Description{Screenshot consisting of an instruction, three items in a table, and a submit button. The instruction reads "Indicate which movies you have watched and if you have liked them." The three items are shown below each other, as three rows. The headings of the table are: "Item," "Seen," "Rating." The item column shows the movie poster and title, seen has two radio buttons with "Yes" and "No" options, and Rating has three radio buttons with "Disliked," "Neutral," and "Liked" options.}
\end{figure*}

\subsection{Explanations}
\label{app:stage2:explanations}

When both positive and negative aspects are displayed (conditions \#3--\#8), we make those fully balanced for each participant as well as across all participants by cycling through the sequences shown in Table~\ref{tab:stage2_sequences}.
These sequences follow a Latin Square design where the binary value P/N is determined by the least significant bit.

\subsection{User Interfaces}
\label{app:ui}

\begin{figure}[t]
    \centering
    \includegraphics[width=0.48\textwidth]{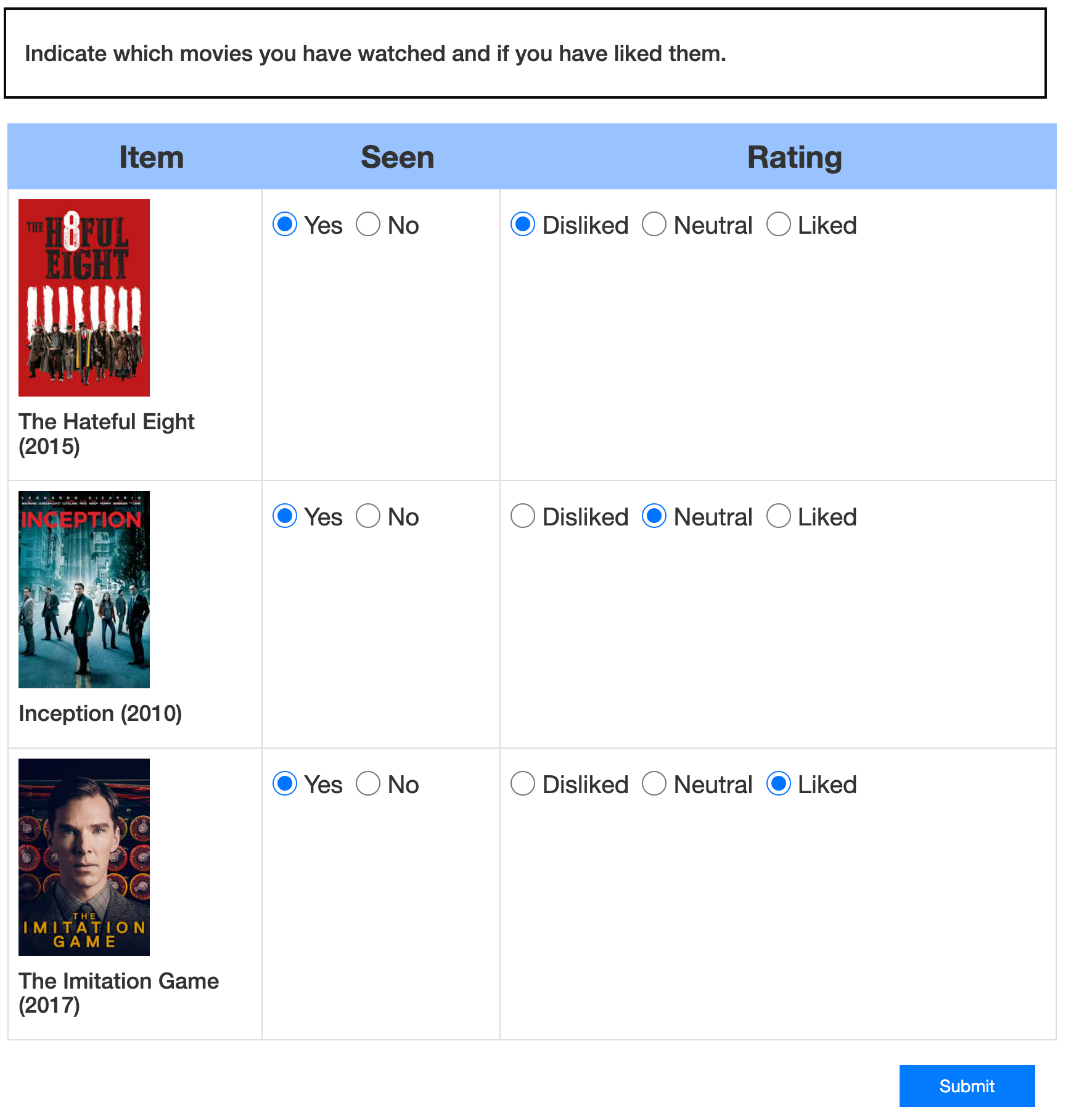}
    \vspace*{-0.5\baselineskip}
    \caption{User interface for item consumption and preference elicitation (Stage 1).}
    \label{fig:task1}
    \Description{Screenshot consisting of an instruction, three items in a table, and a submit button. The instruction reads "Select the movie that you would most likely watch among these." The three items are shown next to each other, as three columns. Each has a movie poster, a title with a radio button in front, a synopsis text, and a list of itemized explanations.}
\end{figure}

Figures~\ref{fig:task1} and~\ref{fig:task2} show screenshots of the user interfaces used in Stages 1 and 2, respectively.

\end{document}